\documentclass[12pt]{article}

\usepackage{amssymb,amsmath,amsfonts,eurosym,geometry,ulem,graphicx,caption,color,setspace,sectsty,comment,footmisc,caption,natbib,pdflscape,subfigure,array,hyperref,circuitikz,tikz,threeparttable,booktabs,tabularx,cleveref,enumitem,setspace,cite,natbib,adjustbox,newtxtext,newtxmath}
\usepackage[title]{appendix}
\setcitestyle{authoryear,round}
\newtheorem{theorem}{Theorem}

\newtheorem{hyp}{Hypothesis}

\newcolumntype{L}[1]{>{\raggedright\let\newline\\arraybackslash\hspace{0pt}}m{#1}}
\newcolumntype{C}[1]{>{\centering\let\newline\\arraybackslash\hspace{0pt}}m{#1}}
\newcolumntype{R}[1]{>{\raggedleft\let\newline\\arraybackslash\hspace{0pt}}m{#1}}

\begin{document}
	\begin{titlepage}
		\title{Import competition and domestic vertical integration: Theory and Evidence from Chinese firms}
		\author{Xin Du \thanks{hhkaaen@hotmail.com; \textit{School of Economics, Wuhan Textile University, Wuhan, China}; ORCID: https://orcid.org/0009-0009-6687-4193}\and Xiaoxia Shi \thanks{corresponding author; 3039440487@qq.com; \textit{School of Economics, Wuhan Textile University, Wuhan, China}} }
		\date{\today}
		\maketitle
		\begin{abstract}
			\noindent What impact does import competition have on firms’ production organizational choices? Existing literature has predominantly focused on the relationship between import competition and firms' global production networks, with less attention given to domestic. We first develop a Nash-bargaining model to guide our empirical analysis, then utilize tariff changes as an exogenous shock to test our theoretical hypotheses using a database of Chinese listed firms from 2000 to 2023. Our findings indicate that a decrease in downstream tariffs lead to an increase in vertical integration. In our mechanism tests, we discover that a reduction in upstream tariffs also enhances this effect. Moreover, the impact of tariff reductions on vertical integration is primarily observed in industries with high asset specificity, indicating that asset-specificity is a crucial mechanism. We further explore whether import competition encourages vertical integration for technological acquisition purpose, the effect is found only among high-tech firms, while it's absent in non-high-tech firms. Our research provides new perspectives and evidence on how firms optimize their production organization in the process of globalization.\\
			\vspace{0in}
			\bigskip
		\end{abstract}
		\setcounter{page}{0}
		\thispagestyle{empty}
	\end{titlepage}
	\pagebreak \newpage
	\onehalfspacing{}
	
	\section{Introduction}
\begin{figure}[htbp]
	\centering
	\captionsetup{justification=centering,skip=0pt}
	\includegraphics[width=0.8\textwidth]{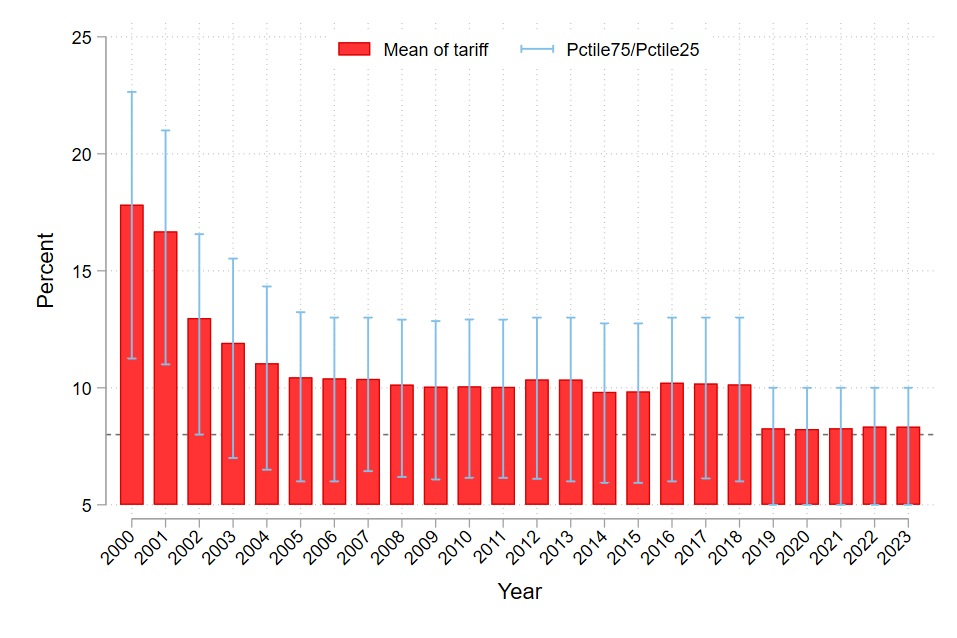}
	\caption{Import tariff change of China between 2000-2023}
	\label{fig_tariff}
\end{figure}

\noindent
	Over the past two decades, trade liberalization has become the foreign trade policy among many developing countries, including China. As shown in \Cref{fig_tariff}, after China's accession to the WTO in 2001, the average tariff level has gradually declined from about 18\% in 2000 to about 8\% in 2023, and lower trade barriers have contributed to the rapid development of global value chains. A large body of literature has explored how trade liberalization relates to the firms' global production network \citep{chorGrowingChinaFirm2021,costinotElementaryTheoryGlobal2013,fernandesDeterminantsGlobalValue2022,bernardTwoSidedHeterogeneityTrade2018}, but less focused given to the firms' production adjustments within national borders \citep{liuTradeLiberalizationDomestic2019,stiebaleImportCompetitionVertical2022}. As shown in \Cref{fig_ma}, the number of domestic mergers and acquisitions (M\&As) by listed Chinese firms has been on a clear upward trend in the past two decades, and vertical M\&As has accounted for most of them. In the context of globalization, firms are compelled to reevaluate not only their global strategies but also the structure of their domestic production organizations. This study contributes to our understanding of both the determinants of firms' production organization choices and the economic consequences of trade liberalization. Moreover, it provides valuable insights to assist governments in developing countries in formulating more effective industrial policies.\par
	We first develop a theoretical model to illustrate the impact of tariff on firms' production organizational choices, provide a theoretical basis for the subsequent empirical analysis. In the model, we assume there are two domestic firms, a Supplier $S$ and a Producer $B$. $S$ supply specialized inputs to $B$, with foreign suppliers provide standardized inputs. Producer $B$ has three choices, buying inputs from $S$ (domestic outsourcing), importing inputs from abroad (offshoring) and integrating with $S$ (domestic integration). Assuming that both $S$ and $B$ aim to maximize profits, first $S$ and $B$ derive the price $P$ of specialized inputs through a Nash-bargaining (if $S$ and $B$ choose to integrate, then there is no price $P$), and then compare the expected profits of the above three options. Our model predicts that a fall in downstream tariffs makes firms more inclined to buy foreign inputs, reduces the level of relationship-specific investment by domestic suppliers, increases the expect profit of integration for buyers, thus induces more vertical integration; a fall in upstream tariffs also result in more vertical integration, and this effect is more primarily observed in differentiated industries, indicates that relationship-specific investment is a key mechanism.\par
	We then selected data on Chinese listed firms from 2000 to 2023 to validate our findings. We use China's import tariff changes as an exogenous shock, measuring the degree of import competition, and the number of vertical M\&As to representing the level of vertical integration. Our empirical results show that tariff reduction has a significant positive effect on backward vertical integration, and passes a series of robustness checks.\par
	Subsequently, we investigate the heterogeneous effects of tariffs on vertical integration to deepen our understanding of existing theories and discuss the underlying mechanisms to test our hypotheses. First, as \citet{fanInstitutionalDeterminantsVertical2017} point out, institutional quality is an important factor influencing vertical integration. we categorize firms into state-owned enterprises (SOEs) and non-SOEs, finding tariff decline significantly promote vertical integration among non-SOEs. Furthermore, firms at different positions in the production chain may exhibit varying behaviors \citep{chorGrowingChinaFirm2021}, we categorize the sample into four quartiles (Q1 lowest, Q4 highest) based on industry upstreamness. The results indicate that tariff reductions significantly impact vertical integration in Q1 and Q3. In the mechanism test, following the methodology of \citet{liuIntermediateInputImports2016}, we construct upstream tariff levels for each industry using Chinese input-output tables and the WITS tariff database. After controlling for upstream tariffs, downstream tariff reductions still promote integration of downstream firms towards upstream sectors, albeit with a decrease in both coefficient magnitude and significance level. Simultaneously, the coefficient for upstream tariff is also negative. This implies that upstream tariff reductions also promote backward integration, consistent with our theoretical hypotheses. Additionally, following the approach of \citet{liuTradeLiberalizationDomestic2019}, we categorize our sample into high and low asset specificity industries to verify the mechanism of relationship-specific investment, the results show that tariff reduction have a significant effect on vertical integration in high asset specificity industries but not in low asset specificity ones, indicating that under-investment of relationship-specific asset serves as another crucial mechanism through which import competition influences vertical integration.\par
	We further examine whether import competition encourages vertical integration for technological acquisition purpose. Technology has long been a crucial driver of vertical integration \citep{acemogluVerticalIntegrationTechnology2010,chunCrossBorderVerticalIntegration2020}. Products with more complex technologies typically require more sophisticated production processes, involving more inter-firm interactions and thus higher transaction costs. this raises the expected benefits of vertical integration. Moreover, due to the characteristics of technological R\&D, such as long development cycles, minimal short-term returns, and high uncertainty, it can be viewed as a form of relationship-specific investment. We hypothesize that intensified import competition may encourage firms to pursue vertical integration for the purpose of technology acquisition.\par
	To verify this mechanism, we first divide our sample into high-tech and non-high-tech firms. We find that the impact of tariff reductions on vertical integration is more pronounced in the high-tech firm sample. Secondly, we separate acquisitions derived by technology acquisition. The results show that increased import competition promotes backward acquisitions for technological purpose among high-tech firms, but this effect is absent in non-high-tech firms.\par
	Our paper makes three main contributions to the existing literature. First, we enriches the relevant research on the economic consequences of import competition. Existing research on the economic consequences of import competition mainly focus on firm performance, such as innovation \citep{liuIntermediateInputImports2016,liuImportCompetitionFirm2021}, product quality \citep{amitiImportCompetitionQuality2013}, mark-ups \citep{luTradeLiberalizationMarkup2015}, productivity \citep{brandtWTOAccessionPerformance2017, topalovaTradeLiberalizationFirm2011}. Some studies focus on the impact of trade liberalization on firms' product scope adjustment \citep{qiuMultiproductFirmsScope2013}, but more on mixed impacts without distinguishing between horizontal and vertical. Relatively less literature has examined the impact of import competition on vertical integration \citep{mclarenGlobalizationVerticalStructure2000,ornelasTradeLiberalizationOutsourcing2008a, ornelasProtectionInternationalSourcing2012}, which explains the phenomenon mainly from a theoretical perspective, but there are few empirical studies to test it. \citet{stiebaleImportCompetitionVertical2022} using Indian data to study the impact of import competition on firms' vertical integration and finds that lower output tariffs induce more vertical integration, while lower input tariffs inhibit vertical integration; \citet{liuTradeLiberalizationDomestic2019} uses data from China to examine the relationship between trade liberalization and firms' vertical integration focusing on upstream and industry-level analysis but ignoring tariffs in the acquirer industries. Our paper extends these studies by investigating the impact of tariff reduction on vertical integration at firm-level and focus on the acquirer industries, thereby complementing existing research. Our approach provides a more comprehensive understanding of how trade liberalization affects firms' boundaries and organizational choices.\par
	Second, our paper contributes to the relevant research on the determinants of vertical integration. There are two main theories explaining vertical integration, the transaction cost economics \citep{williamsonTransactionCostEconomicsGovernance1979} and the property rights theory \citep{grossmanCostsBenefitsOwnership1986,hartPropertyRightsNature1990}. These theories provide a bases for the later research, emphasizing the role of factors such as asset specificity and incomplete contracts in determining firm boundaries. Scholars then shift focus on external factors, \citet{acemogluDeterminantsVerticalIntegration2009} using data from 93 countries, found that both the level of financial development and the cost of contracting significantly contribute to the level of integration. \citet{alfaroPricesDetermineVertical2016} uses tariff changes, an exogenous shock, and finds that higher tariffs raise factor prices, which causes more vertical integration. But cross-country data can omit non-institutional factors such as culture, history or language. Some studies focus on a single country data, \citet{fanInstitutionalDeterminantsVertical2017} uses data on listed firms from China to examine the impact of institutional factors on vertical integration, such as the legal system, market power, and political connections. \citet{acemogluVerticalIntegrationTechnology2010} uses firm data from the UK to study the impact of technology on vertical integration. However, limited literature has focused on the impact of import competition on firms' production organization choices. Our study addresses this gap by examining how import competition influences vertical integration, also provide some insights on which channel through import competition influences vertical integration, contributing to a more comprehensive understanding of the determinants of firm boundaries in the context of international trade.\par
	Third, we provide new empirical evidence and theoretical insights on how firms in developing countries respond to trade openness. As the world's second-largest economy and the leading developing country, China offers a unique context for studying the challenges and opportunities faced by firms in emerging markets during the process of globalization, provide insights for policymakers in formulating industrial policies.\par
	The rest of the paper is organized as follows: in Section 2 we present a theoretical model to explore how tariff affects firms' production choices; Section 3 explains the data and empirical strategy; Section 4 reports the empirical results; and finally, Section 5 concludes.	
	
	\section{Theoretical Framework}\noindent
	We build a Nash-bargaining model based on \citet{ornelasTradeLiberalizationOutsourcing2008a,ornelasProtectionInternationalSourcing2012} and \citet{ liuTradeLiberalizationDomestic2019} to have a better understanding on how tariff affect firms' production organizational choices.
	
	\subsection{Basic model}\noindent
	Suppose that there are two domestic firms, an upstream supplier $S$ and a downstream producer $B$. The domestic supplier $S$ has an information advantage compared to the foreign suppliers, thus can produce specialized inputs, while the foreign supplier only produces standardized inputs. The price of the specialized inputs $P$ is decided by $S$ bargaining with $B$, while the price of the standardized inputs, $p_f$, is determined by the foreign market and is randomly distributed in $[0,\infty]$, the distribution function is $F(p_f)$. Since the foreign supplier only produces standardized inputs, there is no ``hold-up problem" between the foreign supplier and the domestic producer, so it makes no difference between offshoring and cross-border integration for $B$. Under this circumstance, $B$ has three options: 1. purchase inputs from domestic supplier $S$ (domestic outsourcing); 2. import standardized inputs from abroad (offshoring); and 3. integrate with domestic supplier $S$ (integration).\par
	When $B$ buys inputs from domestic supplier $S$, the price of specialized input $P$ is derive from Nash-bargaining, we assume that the bargaining power is $\alpha$ and $(1-\alpha)$ for $S$ and $B$; if standardized inputs are purchased from foreign supplier, they need to pay price $p_f$ with input tariff $\tau$, we call this the \textit{upstream tariff} (assuming $\tau$ is not prohibitive); if $B$ chooses to integrate with domestic suppliers, it pays an additional fixed cost $K$, as shown in \Cref{relation}.\par
	Wherever $B$ purchases inputs from, its value is $V$, affected by the tariff $t$, we call this the \textit{downstream tariff}, we have $V=V(t)$, with $V'(t)<0, V''(t)<0$, i.e., the value from input purchases by $B$ diminishes at the margin with the tariff. For firm $S$, the cost of producing specialized inputs is $c$, meantime supplier $S$'s \textit{ex-ante} investment in specialized input is $e$, and we assume $c=c(e)$, with $c'(e)<0, c''(e)<0$ (the more $S$ invests, the less it costs to produce the specialized input). $B$ will buy inputs from $S$ if $c(e)\leq \min(p_{f}+\tau,V(t))$, but since $p_f$ is a random distributed variable, we assume $c(e)\leq V(t)$, which makes the expected profit of buyer $B$ is positive regardless of its choice.\par
	The timing of the game is as follows: at time 1, the buyer and the seller decide whether to integrate or not. In time 2, the seller $S$ chooses an optimal investment level if they do not integrate, or they decide an optimal investment level after integration if they decide to integrate. At time 3, the price of foreign inputs $p_f$ is determined. At time 4, if the buyer and seller do not integrate at time 1, then $B$ choose to buy input whether from domestic or abroad; if two parties integrate at time 1, the integrated firm chooses between produce input within-firm and import from abroad. At time 5, production and trade occur (if at all). Note that buyer $B$ cannot commit ex ante (i.e., before $p_f$ is realized) to buy either domestic or foreign inputs, but the buyer will make an efficient decision \textit{ex post}.
	
	\subsection{Non-integration}\noindent
	Suppose that at time 1, the buyer and seller do not to integrate. Then the buyer bargaining with the seller. If the bargaining is successful , they agree on a price $P$, then the buyer's expected profit is $u_b^1=V(t)-P$ and the seller's profit is $u_s^1=P-c(e)$; if the bargaining fails, price $P$ fails to be formed, and there is no profit to the seller, and the buyer imports inputs from abroad, then the buyer's profit is $u_b^0=\max(V( t)-p_f-\tau,0)$, and the seller's profit is $u_s^0=0$. We assume that successful bargaining is a dominant decision for both parties, with $u_b^1>u_b^0,u_s^1>u_s^0$. The price $P$ is the solution to the Nash-bargaining given by the following equation:
	\begin{equation}
		P(\tau)=\mathop{\arg\max}\limits_{P}(u_{s}^{1}-u_{s}^{0})^{\alpha}(u_{b}^{1}-u_{b}^{0})^{1-\alpha}
	\end{equation}
	Deriving this equation for $P$, we have the distribution of domestic input price $P$:
	\begin{equation}
		P(\tau)=\begin{cases}
			\alpha(p_f+\tau)+(1-\alpha)c(e) & \text{if $V(t)\geq p_f+\tau$}\\
			\alpha V(t)+(1-\alpha)c(e) & \text{if $V(t)< p_f+\tau$}
		\end{cases}
	\end{equation}\par
	At time 2, according to price $P$, the seller chooses an optimal investment level $e$ to maximize its profit. Since $p_f$ is a randomly distributed variable, we need to discuss it in three cases:
	\begin{equation}\label{situation}
		\begin{cases}
			p_f+\tau \leq c(e) \leq V(t) & \textcircled{1}\\
			c(e)< p_f+\tau \leq V(t) & \textcircled{2}\\
			c(e)\leq V(t) < p_f+\tau & \textcircled{3}
		\end{cases} 
	\end{equation}\par
	When case \textcircled{1} occurs, the buyer imports inputs from abroad, so the profit of $S$ is $u_s=-e$; when case \textcircled{2} occurs, the buyer imports inputs from domestic, we have $u_s=P-c(e)-e$; when case \textcircled{3} occurs, the buyer is still importing inputs from domestic, but at different price $P$, we have $u_s=P-c(e)-e$. In total, the expect profit of the seller $S$ is:
	\begin{equation}
		u_s(e,t,\tau)=\int_{c(e)-\tau}^{V(t)-\tau}\alpha(p_f+\tau-c(e))dF(p_f)+\int_{V(t)-\tau}^{\infty}\alpha(V(t)-c(e))dF(p_f)-e
	\end{equation}\par
	Assume that $\frac{\partial^2 u_s}{\partial e^2}<0$, so there is only one optimal investment level $e^n$, leads to $\frac{\partial u_s}{\partial e}=0$, $e^n$ derives from $-c'(e^n)[1-F(c(e^n)-\tau)]=\frac {1}{\alpha}$. Thus, the expected profit of the seller is $u_s(e^n(t,\tau),t,\tau)$, and the expected profit of the buyer is $u_b(e^n(t,\tau),t,\tau)$, and sum these two together we can have the total expected profit for non-integration as $U^n(t,\tau)=u_s(e^n(t,\tau),t,\ tau)+u_b(e^n(t,\tau),t,\tau)$.
	
	\subsection{Vertical Integration}\noindent
	Assuming that at time 1, both parties choose to integrate, then there is no price $P$, and the integration firm chooses between producing inputs within-firm or importing from abroad. Since $p_f$ is a random distributed variable, we also need to discuss the three cases in \cref{situation}: when case \textcircled{1} occurs, the integration firm chooses to import from abroad, the expect profit would be $u_v=V(t)-p_f-\tau-e$; when case \textcircled{2} occurs, the integration firm produces inputs within-firm, so the expect profit is $u_v=V(t)-c(e)-e$; when case \textcircled{3} occurs, the integration firm's expect profit is the same as case \textcircled{2}. Then sum the above three cases, we have the expect profit for the integration firm as:
	\begin{equation} \label{eq4}
		u_v(e,t,\tau)=\int_{0}^{c(e)-\tau}(V(t)-p_f-\tau)dF(p_f)+\int_{c(e)-\tau}^{\infty}[V(t)-c(e)]dF(p_f)-e
	\end{equation}\par
	We assume that there is also only one optimal investment level $e^v$ in the integration case, obtained from $-c'(e^v)[1-F(c(e^v)-\tau)]=1$. Bringing $e^v$ into \cref{eq4} we get the expected profit of the integration firm as $U^v(t,\tau)=u_v(e^v(t,\tau),t,\tau)$. Comparing the conditions for determining the level of investment $e$ at non-integration and integration, we can conclude:
	\begin{theorem}
		The optimal investment level under vertical integration is greater than under non-integration, i.e., $e^v>e^n$.
	\end{theorem}
	
	\subsection{Organizational Choice}\noindent
	Suppose there is no wealth constraint and all agents are risk neutral. Integration will only occur when the difference between post-integration profits and non-integration profits exceeds $K$. Our focus is on how changes in downstream tariff $t$ and the upstream tariff $\tau$ affect the profit differential in $\Delta U$ (where $\Delta U=U^v-U^n$). By taking partial derivatives of $\Delta U$ with respect to $t$ and $\tau$, we can determine how downstream and upstream tariffs respectively influence firms' production choices. Specifically, we derive $\frac{\partial \Delta U(t,\tau)}{\partial t}$ and $\frac{\partial \Delta U (t,\tau)}{\partial \tau}$. Given that $\Delta U=U_{v}-U_{n}$, we can expand $\frac{\partial \Delta U(t,\tau)}{\partial t}$ as follows:
	\begin{align}\notag
		\frac{\partial \Delta U}{\partial t} 
		&=\frac{\partial U^v}{\partial t}-\frac{\partial U^n}{\partial t}\\ \notag
		&=\left( \frac{\partial U^v}{\partial e^v}\cdot \frac{\partial e^v}{\partial t}+\frac{\partial U^v}{\partial t}\right) -\left( \frac{\partial U^n}{\partial e^n}\cdot\frac{\partial e^n}{\partial t}+\frac{\partial U^n}{\partial t}\right) \\
		&=\left( \frac{\partial U^v}{\partial e^v}\cdot \frac{\partial e^v}{\partial t}+\frac{\partial U^v}{\partial t}\right) -\left( \frac{\partial (u_s+u_b)}{\partial e^n}\cdot\frac{\partial e^n}{\partial t}+\frac{\partial U^n}{\partial t}\right)
	\end{align}\par
	Given that $e^{v}$ represents the optimal investment level under integration, and $e^{n}$ represents the optimal investment level under non-integration, we have $\frac{\partial u_s}{\partial e^n}=0$ and $\frac{\partial U_v}{\partial e^v}=0$. Consequently, the above expression can be further simplified to:
	\begin{align}\label{eq6}
		\frac{\partial \Delta U(t,\tau)}{\partial t} 
		&=\frac{\partial U^v}{\partial t}-\frac{\partial U^n}{\partial t}-\frac{\partial u_b}{\partial e^n}\cdot\frac{\partial e^n}{\partial t}
	\end{align}\par
	Following the same logic above, we can derive:
	\begin{align}\label{eq7}
		\frac{\partial \Delta U(t,\tau)}{\partial\tau} =\frac{\partial U^v}{\partial\tau}-\frac{\partial U^n}{\partial\tau}-\frac{\partial u_b}{\partial e^n}\cdot\frac{\partial e^n}{\partial \tau}
	\end{align}\par
	\Cref{eq6,eq7} demonstrate that the impact of tariff changes on the profit differential between integration and non-integration can be decomposed into three terms. The first term represents the marginal effect of tariff changes on the total profits of integration firm. The second term captures the marginal effect of tariff changes on the total profits of non-integration firm. The third term accounts for the indirect effect of tariffs through changes in investment levels. Simplify \cref{eq6,eq7}, we ultimately obtain:
	\begin{equation}\label{key1}
		\frac{\partial \Delta U(t,\tau)}{\partial t}=-\frac{1-\alpha}{\alpha}\cdot \frac{\partial e^n}{\partial t}
	\end{equation}
	\begin{equation}\label{key2}
		\frac{\partial \Delta U(t,\tau)}{\partial \tau}=F(c(e^n)-\tau)-F(c(e^v)-\tau)-\frac{1-\alpha}{\alpha}\cdot \frac{\partial e^n}{\partial \tau}
	\end{equation}\par
	First, we discuss the marginal impact of changes in downstream tariff $t$. In \cref{key1}, ceteris paribus, when the downstream tariff decreases, the direct effect on firms is identical whether they are integrated or non-integrated, thus these effects cancel each other out. However, a reduction in $t$ increases the buyer's profits from purchasing inputs. Under similar conditions, buyer $B$ always tend to imported inputs from abroad. When the domestic seller $S$ anticipates this, they will reduce the relationship-specific investment $e$ ($\frac{\partial e^n}{\partial t}>0$), exacerbating the under-investment problem. When the under-investment problem become more severe, the benefits for buyer $B$ to integrate increase. Consequently, a reduction in tariff $t$ promotes more integration. Based on the above analysis, we propose the following hypothesis:
	\begin{hyp}\label{propo1}
		A reduction in downstream tariff $t$ will induce more vertical integration.
	\end{hyp}\par
	Next, we discuss the marginal impact of upstream tariff $\tau$. In \cref{key2}, there are two forces at work. On the one hand, when upstream tariff $\tau$ decreases, the buyer's demand shifts from domestic to abroad, foreign outsourcing become more attractive, which reduces the benefits of integration for the buyer, encouraging more foreign outsourcing but less domestic integration, that is $F(c(e^n)-\tau)-F(c(e^v)-\tau)>0$. On the other hand, lower upstream tariffs shift domestic buyers' demand from domestic inputs to foreign inputs, this leads domestic sellers to decrease relationship-specific investment. Thus, a decrease in upstream tariff $\tau$ result in reduced investment by domestic seller, i.e., $\frac{1-\alpha}{\alpha}\cdot\frac{\partial e^n}{\partial \tau}>0$. However, as domestic sellers reduce their investment, the benefits of integration for domestic buyers increase. If $e^n$ is highly responsive to $\tau$ (e.g., reaching a critical value) and $c(e)$ is sufficiently concave, the effect in the second part of \cref{key2} may dominate the first part. Ultimately, a decrease in upstream tariff $\tau$ may promote more integration, i.e., $\frac{\partial \Delta U(t,\tau)}{\partial \tau}<0$. Based on the above analysis, we propose our second hypothesis:
	\begin{hyp}\label{propo2}
		A reduction in upstream tariff $\tau$ will induce more vertical integration.
	\end{hyp}\par
	\Cref{key1,key2} also reveals that the bargaining power of seller and buyer $(\alpha,1-\alpha)$ influences our conclusions. In low asset-specificity industries, products are highly standardized, allowing suppliers and producers to match randomly. Due to low differentiation between products, the demand for relationship-specific investments is relatively low, resulting in more balanced bargaining power between downstream and upstream. Conversely, in high asset-specificity industries, product differentiation is significant high, leading to a higher demand for relationship-specific investments. This creates a larger disparity in bargaining power between buyers and sellers. In buyer-dominated markets, where $\alpha$ is relatively small and ($1-\alpha$) is large, the marginal effects of tariffs are amplified. Thus we propose our third hypothesis:
	\begin{hyp}\label{propo3}
		The effect of tariff reduction on vertical integration is significantly larger in high asset-specificity industries relative to low industries.
	\end{hyp}\par
	Our model differs from existing literature in two main aspects. Firstly, our focus is on downstream rather than upstream. Previous studies on the impact of trade liberalization on vertical integration have primarily focused on the upstream tariffs. \citet{ornelasTradeLiberalizationOutsourcing2008a} found that a decrease in intermediate tariffs leads to an increase in vertical integration. \citet{ornelasProtectionInternationalSourcing2012} further discovered that tariffs can improve social welfare under certain conditions by addressing specific market failures (such as the "hold-up problem"), but their focus remained on intermediate tariffs. \citet{liuTradeLiberalizationDomestic2019} simultaneously examined the effects of reductions in both input and output tariffs for upstream firms on vertical integration. Building upon these studies, we focus on downstream, investigating the impacts of both input tariffs (i.e., upstream tariffs) and output tariffs (i.e., downstream tariffs) on vertical integration.\par
	Secondly, we focus solely on domestic vertical integration. Cross-border M\&A typically involve more complex influencing factors compared to domestic acquisitions. Moreover, in many countries, domestic acquisitions often dominate, making the focus on domestic acquisitions more significant. The rest of the paper uses data from Chinese listed firms from 2000 to 2023 to test our Hypotheses.
	\section{Empirical method and Data}\noindent
	\subsection{Empirical method}\noindent
	We use Chinese listed firms from 2000 to 2023 to identify the impact of import competition on vertical integration, our econometric model is specified as follows:
	\begin{equation}\label{baseline_eq}
		\ln(backward_{fit})=\beta\cdot tariff_{it}+X_{fit}'\Gamma+\delta_f+\delta_i+\delta_t+\varepsilon_{fit}
	\end{equation}
	In \cref{baseline_eq}, the dependent variable $backward_{fit}$ represents the number of backward acquisitions made by firm $f$ in industry $i$ during year $t$. The classification of acquisitions as backward or forward is based on the upstreamness of the industry. The key dependent variable $tariff_{it}$ denotes the level of import tariffs in industry $i$ during year $t$. $X_{fit}$ is a vector of firm- and industry-level control variables. Fixed effects are denoted by $\delta$, subscripts $f$, $i$, and $t$ denote firm, industry, and year respectively, and the error term is denoted by $\varepsilon_{fit}$. Since the dependent variable is the number of M\&As, we use Poisson Pseudo Maximum Likelihood (PPML) Estimation to test our theoretical hypotheses.
	\subsection{Definition of key variables}\noindent
	To define whether an M\&A event is backward, we refer to the industry's upstreamness calculated by \citet{chorGrowingChinaFirm2021} to define the position of a firm in the whole production line, calculated as follows:
	\begin{equation}
		UPS_i=1\cdot \frac{F_i}{Y_{i}}+2\cdot \frac{\sum_{j=1}^{N}d_{ij}F_{j}}{Y_{i}}+3\cdot \frac{\sum_{j=1}^{N}\sum_{k=1}^{N}d_{ik}d_{kj}F_{j}}{Y_{i}}+\ldots,
	\end{equation}
	where $Y_{i}$ denote the total output of industry $i$, $F_{i}$ is the value of output goes directly to final investment or consumption, $d_{ij}$ is the direct requirement coefficient in the Chinese IO tables, denotes  the value of input from industry $i$ needed to produce one unit of output in industry $j$. Then we manually match the Chinese IO tables to the China's Industrial Classification (CIC), along with conversion table between CIC and the International Standard Industrial Classification (ISIC) provided by the National Bureau of Statistics (NBS), we ultimately adopt the upstreamness at four-digit ISIC level.\par
	Our M\&As data comes from Bureau van Dijk (BvD)'s zephyr Global M\&A Database. We select data based on the following steps: 1. Deal type as Acquisition\&Merger; 2. Deal status as Announced\&Completed; 3. Time period from January 1, 2000, to January 1, 2024; 4. The location of the firm is China; 5. Company status as listed acquirer. Since we only focus on domestic vertical integration, we retain only firms in the mainland. Combined with the upstreamness at the four-digit ISIC level, we define: In an acquisition event, if the upstreamness of the acquiring firm differs from that of the acquired firm, the event is classified as vertical integration. Further, we compare the upstreamness of the acquiring and acquired firms. If the upstreamness of the acquired firm is higher than that of the acquiring firm, the event is defined as backward integration ($backward$).\par
	We use Most Favored Nation (MFN) tariffs represent the degree of import competition. Tariff data is from the WITS database and the Tariff Download Facility database. MFN tariffs are agreed upon after long rounds of multilateral trade negotiations, at the end of each round, governments commit not to exceed the agreed tariff rate. In the event of non-compliance with this commitment, the affected parties have the right to refer the matter to the WTO Dispute Settlement Body. Once agreed, tariff rates must be applied in a non-discriminatory manner to imports from all WTO members \citep{alfaroPricesDetermineVertical2016}, severely limits negotiators’ flexibility to respond to lobbying, which provides us an exogenous shock.\par
	We next combine the tariff data and the Chinese IO-tables to calculate the upstream tariffs for each industry. Following \citet{liuIntermediateInputImports2016}, we calculate the upstream linkages of each 4-digit CIC industry using the complete requirement coefficients. This approach allows us to compute the upstream tariff level for each industry $i$ in year $t$. Based on the transformation codes of the Chinese IO Tables, we first convert the HS 6-digit level tariffs to IO sectors, represented in the matrix as:
	\[
	tariff_{t,j} = \left( {\begin{array}{*{20}{c}}
			{tarif{f_{2000,1}}}&{tarif{f_{2000,2}}}&{...}&{tarif{f_{2000,j}}} \\ 
			{tarif{f_{2001,1}}}& \ddots &{}&{tarif{f_{2001,j}}} \\ 
			\vdots &{}& \ddots & \vdots  \\ 
			{tarif{f_{t,1}}}&{tarif{f_{t,2}}}&{...}&{tarif{f_{t,j}}} 
	\end{array}} \right)
	\] \par
	Where $t$ is year, $j$ is IO-sector. Each row represents the output tariff of all sectors in year $t$; each column represents the output tariff of all years in industry $j$. And the matrix of direct requirement coefficients is:
	\[
	IO_{j,i} = \left( {\begin{array}{*{20}{c}}
			{{x_{1,1}}}&{{x_{1,2}}}& \cdots &{{x_{1,i}}} \\ 
			{{x_{2,1}}}& \ddots &{}&{{x_{2,i}}} \\ 
			\vdots &{}& \ddots & \vdots  \\ 
			{{x_{j,1}}}&{{x_{j,2}}}& \cdots &{{x_{j,i}}} 
	\end{array}} \right)
	\] 
	Where $IO_{j,i}$ denotes the value of input from industry $j$ needed to produce one unit of output in industry $i$, then combine the two matrix above, we have:
	\begin{equation}\label{matrix}
	upstream\_tariff_{t,i}=tariff_{t,j}\cdot IO_{j,i}=\left( \begin{matrix}
		input_{2000,1} & \ldots & input_{2000,i}\\
		\vdots & \ddots  & \vdots\\
		input_{t,1} & \cdots & input_{t,i}\\
	\end{matrix}\right)
	\end{equation}\par
	In \cref{matrix},the rows of the matrix denote the level of upstream tariffs of all sectors in year $t$, and the columns denote the level of upstream tariffs of all years in sector $i$. Through the conversion table mentioned above, we obtain the upstream tariffs for each 4-digit ISIC industry $i$.\par
	The control variables are from the WIND database of listed firms. We select six indicators as our control variables: firm age ($age$), firm size ($size$), asset liquidity ($liquidity$), leverage ratio ($leverage$), and HHI index ($hhi$), the specific calculation methods for these variables are detailed in \Cref{control}. To mitigate the influence of extreme values, we winsorize all continuous variables at the 1st and 99th percentiles. Additionally, we excluded firms with ``ST" and ``*ST" in their names, which indicates financial distress or other irregularities. We also exclude firms in the financial sector, identified by 2-digit CIC codes ``66", ``67", ``68", and ``69". \Cref{describe} are our descriptive statistics.
	
	\section{Emperical results}\noindent
	
	\subsection{Baseline results}\noindent
	To test \Cref{propo1}, we use Poisson Pseudo Maximum Likelihood (PPML) Estimation, the results are shown in \Cref{baseline}. Initially, we excluded fixed effects and control variables, as presented in column (1), a 1\% reduction in tariffs leads to an increase in the expected number of vertical integration by around 2.4\%. Subsequently, we introduced fixed effects, as shown in column (2), the impact remains, and the explanatory power of the model significantly increases. In column (3), after adding control variables, the coefficient remains significantly negative. Finally, we add both control variables and fixed effects, our conclusions still hold. This indicates that when downstream tariffs are reduced, downstream firms are more likely to integrate towards upstream.
	
	\subsection{Robustness checks}\noindent
	We next verify the robustness of the baseline results, consider a range of confounding factors that may affect the results, such as the model setting, the measurement of the variables, etc., and the results still remain robust.
	
 	\subsubsection{Switch models}\noindent
 	We use alternative regression models to test the robustness of our findings. In \Cref{robust1}, we first replace the PPML model with an OLS model, as shown in column (1), the coefficient is significant at the 1\% level, the direction of coefficient remains consistent with the baseline results. It is important to note that the PPML model is more suitable for count data characteristics, thus the coefficients more accurately reflect the true relationship. Though the magnitudes of coefficients have been changed quite significantly, but the significance level and direction remain unchanged, supporting the robustness of our baseline results. Secondly, we use logit model as a robustness check, by generate binary variables indicating whether firm $i$ has vertical integration in year $t$, again distinguishing between forward and backward integration, the result is shown in column (2) of \Cref{robust1}, consistent with the baseline results.
	
	\subsubsection{Switch samples}\noindent
	In the following analysis, we alter the sample size to test the robustness of the baseline results. First, we exclude firms with excessively high leverage ratios, which usually means that the firm is in financial distress, including such firms could potentially bias our results. We exclude firms from the sample whenever a firm's leverage ratio ever exceeds 1 during the sample period, and the result remain robust as shown in column (3) of \Cref{robust1}.\par
	Second, we exclude non-manufacturing firms from our sample. Manufacturing firms usually involve highly specialized assets, such as customized production equipment and molds. Reduction in tariffs would exacerbate the problem of under-investment in the upstream, increase the benefits of vertical integration in downstream firms, induces more vertical integration. However, in non-manufacturing sectors (e.g., service sector, agriculture), asset specificity may reflect in specialized skills or intellectual property, but tariffs are usually taxes on tangible goods, thus the impact of tariffs is limited for non-manufacturing firms. As shown in column (4) of \Cref{robust1}, remaining robust.\par
	Additionally, we exclude the sample during the financial crisis (2008-2009), since it's a significant external shock event. This may cause firms to behave unconventionally, and these behaviors may interfere with our conclusions. We exclude data from 2008 to 2009, result is shown in column (5) of \Cref{robust1}, the magnitude and the significance level of the coefficients is not much affected by this. Finally, we use all the above methods in column (6), and the result are still robust.
	
	\subsubsection{Alter the measurement method}\noindent
	We then alter the measurement method of dependent variable to test the robustness of our baseline results. First, we replace MFN tariff with AHS tariff. Under WTO rules, the MFN tariff rate must be applied in a non-discriminatory manner to imports from all WTO members, but there are exceptions (e.g. regional free trade agreements). But AHS tariff is the actual tariff rate that a country imposes on imports, including circumstances like trade agreements, special market access conditions, or preferential programs, which may lead to lower tariff than MFN, hence represent a closer situation of real tariff.
	The result is shown in column (1) of \Cref{robust2}, continue to support our conclusions.\par
	Second, we adjust the tariff from 4-digit ISIC level to 3-digit level. This is  mainly based on two concerns: first, the production activities of some firms may span multiple industries, 
	thus make 4-digit tariff not enough to reflecting the competition faced by these firms \citep{liuTradeLiberalizationDomestic2019}; and second, fixed effects at the 4-digit industry level may be insufficient to accurately capture the intrinsic industry trend, thereby confuse the industry trend with the effect of import competition. The result is shown in column (2) of \Cref{robust2}, the coefficients of vertical and backward remain largely unchanged, but the effect on forward also turn to significant, since our primary focus is on backward, this adjustment does not compromise the robustness of our conclusions.\par
	Furthermore, we alter the simple tariffs into weighted tariffs. Tariffs are initially based on the 6-digit HS codes, while our study defines industries using the 4-digit ISIC codes. During the conversion, we use simple average tariffs in our baseline regression to partially circumvent endogeneity concerns, this approach may not adequately reflect the actual tariff levels of industries. Therefore, we employ weighted tariffs as a robustness check, with the specific formula as follows:
	\begin{equation}\label{13}
		value_{i}=\sum_{j=i}^{}value_{j}
	\end{equation}
	First, we aggregate the trade values of products from the 6-digit HS codes to the 4-digit ISIC codes using a conversion table. In \cref{13}, $j$ represents a product under the 6-digit HS code, and $i$ represents a product under the 4-digit ISIC code. $value_j$ denotes the trade volume of HS 6-digit product $j$, while $value_i$ denotes the aggregate trade volume of the ISIC 4-digit product $i$. Then we have:
	\begin{equation}\label{14}
		weight\_tariff_{i}=\sum_{j=i}(tariff_{j}\times \frac{value_{j}}{value_{i}})
	\end{equation}\par
	In \cref{14}, $tariff_{j}$ represents the tariff level for HS 6-digit product $j$, multiplied by the weight $\frac{value_{j}}{value_{i}}$, which denotes the proportion of trade volume of product $j$ in the total trade volume of the aggregated product $i$. The sum of these weight tariffs yields the weight-average tariff $weight_tariff_{i}$ for product $i$. The regression result are presented in column (3) of \Cref{robust2}, remain significantly negative, thus confirming the robustness of our conclusions.

	\subsubsection{Endogeneity issue}\noindent
	Although tariff change is an exogenous shock, our identification strategy is still endogenous to some extent, such as reverse causality and the problem of omitted variables. First, we lag tariffs by one period to avoid reverse causality. The approach yields two benefits: 1. capturing the lagged effect of tariffs on vertical integration; and 2. avoiding reverse causality to some extent. The result remain robust as shown in column (4) of \Cref{robust2}.\par
	Second, to avoid the problem of omitted variables, we include additional factors that may affect the vertical integration. These factors include: the firm's fixed asset ratio ($tangibility$), denoted by the logarithm of (fixed assets/total assets+1); capital density ($k\_density$), denoted by the logarithm of (fixed assets/number of employees+1); and the growth rate of turnover ($growth$), denoted by the logarithm of (turnover in year $t$/turnover in year $(t-1)$). The result are shown in column (5) of \Cref{robust2}, still robust.
	
	\subsection{Heterogenous effects}\noindent
	We demonstrate the impact of tariff reduction on domestic vertical integration in the baseline results and robustness, we next further explore the heterogeneous effect of tariff reduction.
	
	\subsubsection{Non-soe and soe}\noindent
	First, we divide the sample into SOEs and non-SOEs to examine the heterogeneous effects of ownership, the results are shown in columns (1)-(2) of \Cref{hetero1}. We find that the impact of tariff reductions is significant only among non-SOEs, while insignificant among SOEs. The emergence of this result is not surprising; harsh government regulations restrict private firms' operations \citep{fanInstitutionalDeterminantsVertical2017}, while bureaucrats have the right to bypass these regulations, so SOEs face fewer competitions and lack the motivation for vertical integration.
	
	\subsubsection{Upstreamness, from low to high}\noindent
	Second, we categorize firms into four groups to explore the heterogeneous effects of tariff reduction at different production line positions. Previous literature finds that firms at different positions in the production chain may exhibit significantly different behaviors \citep{chorGrowingChinaFirm2021}. As shown in columns (3)-(6) of \Cref{hetero1}, the firms are ranked according to the upstreamness of industry from low to high (Q1-Q4). Overall, the impact of import competition on backward vertical integration is concentrated in the Q1 (closest to the consumption side) and the Q3 (second closest to the production side). For the lowest firms (Q1), facing direct import competition may incentivize them to enhance competitiveness and control upstream inputs through backward integration. For the second-highest firms (Q3), backward integration may serve to ensure a stable supply of critical inputs or reduce transaction costs with upstream suppliers.
	
	\subsection{Mechanism tests}\noindent
	In our theoretical analysis, we emphasize the differential impact of upstream on firms' incentives to vertical integration, as well as the importance of relationship-specific investments. In this section, we examine the roles of these two factors.
	
	\subsubsection{Upstream tariff}\noindent
	To test \Cref{propo2}, We analyze the change of coefficient on downstream tariff before and after incorporating upstream tariffs, as presented in \Cref{mechanism}. From columns (1)-(2), the magnitude and significance level of the coefficient on backward integration decreases compared to no upstream tariffs are included. Moreover, the coefficient of upstream tariffs is significantly negative, indicating that upstream tariff reductions promote backward integration. To further support this conclusion, we examined the changes in the coefficients of forward integration before and after including upstream tariffs\footnote{Results are not shown in the main text but are available upon request from the authors}. The results show that after incorporating upstream tariffs, the impact of tariff reductions on forward integration becomes significant, whereas it was previously insignificant. These findings support our \Cref{propo2}.
	
	\subsubsection{Relationship-specific investment}\noindent
	We next examine \Cref{propo3}. Industries with high asset specificity require more relationship-specific investment, potentially leading to more ``hold-up problem''. Therefore, we divide our sample into two subsamples: industries with high asset specificity and low asset specificity, comparing the difference in coefficients across subsamples. Following the approach of \citet{liuTradeLiberalizationDomestic2019} and \citet{stiebaleImportCompetitionVertical2022}, we employ the classification method of \citet{rauchNetworksMarketsInternational1999}, categorizes industries into three types: ``differentiated'', ``traded on organized exchanges'', and ``reference priced''. We combine the latter two into a catch-all ``homogeneous'' industries and differentiated goods as ``heterogeneous'' industries. The results are shown in columns (3)-(4) of \Cref{mechanism}. The impact of tariff reductions on backward vertical integration is significant at the 1\% level in heterogeneous industries, while insignificant in homogeneous industries. These results support our \Cref{propo3}, suggesting that tariff reduction can reduce the investment made by upstream, thus increase the expect profit of downstream firms to integrate with upstream, finally induces more vertical integration.

	\subsection{Further discussion: Technological VI}\noindent
	In this section, we further discuss technology as a mechanism that tariff impact firm's organizational choices. Our previous analysis centered on relationship-specific investments, as standardized products, being easily described and valued, are well-suited to arm's-length market transactions. Moreover, the diversity of suppliers and buyers for standard products mitigates issues arising from asset specificity. But transaction costs increase when inter-firm relationships require greater coordination, non-standard inputs involve more complex transfers of design information and therefore intense interactions across firm boundaries\citep{gereffiGovernanceGlobalValue2005}, which induces more vertical integration. However, investment in technology itself is also a form of specific asset. Technological R\&D is characterized by high uncertainty, long cycles, and significant risks. Moreover, products with higher technological content may involve more complex production processes, requiring more inter-firm interactions. This suggests that high-tech industries or firms may face similar underinvestment problems. Consequently, import competition might have heterogeneous effects on these industries or firms.\par
	Existing research has demonstrated the impact of technology on vertical integration. \citet{acemogluVerticalIntegrationTechnology2010} examined the impact of technology on vertical integration using data from the United Kingdom. They found that the higher the technological intensity of downstream firms, the greater the likelihood of vertical integration. Conversely, the higher the technological intensity of upstream firms, the lower the probability of vertical integration. \citet{chunCrossBorderVerticalIntegration2020} discovered a similar pattern among Korean firms. Cross-border vertical integration is more likely when the domestic downstream industry has high R\&D intensity and the foreign upstream industry has low R\&D intensity.
	Based on their work, we further examine whether technology serves as a potential channel through which import competition affects vertical integration.\par
\begin{figure}[htbp]
	\centering
	\captionsetup{justification=centering}
	\includegraphics[width=0.8\textwidth]{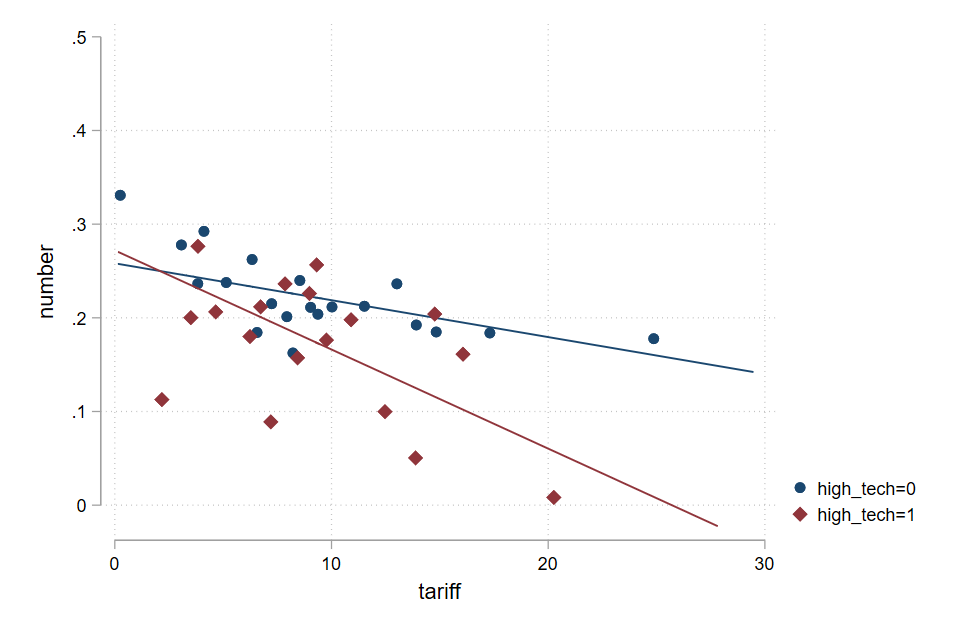}
	\caption{high-tech firms and non high-tech firms}
	\label{fig_hightech}
\end{figure}

	First, we divide the sample into high-tech firms and non-high-tech firms, and run regressions based on \cref{baseline_eq}. The definition of high-tech firms comes from Wind database and the regression results are shown in columns (1)-(2) of \Cref{hightech}. We find the effect of import competition on vertical integration is larger in high-tech firms. Using the methodology provided by \citet{cattaneoBinscatter2024}, we visualize this result as shown in \Cref{fig_hightech}, the diamond-shaped points represent high-tech firms, while circular points represent non-high-tech firms. We find the sloop of high-tech firm is clearly steeper than non-high-tech firms, this aligns with previous studies \citep{ acemogluVerticalIntegrationTechnology2010}.\par
	Second, we categorize M\&As into technology-driven and non-technology driven M\&As to test whether increased import competition prompts firms to engage in more technological M\&As. technological M\&As refers to whether an acquisition's purpose is to obtain technological assets such as technology, patents, research and development capabilities, etc. Technological M\&As can help firms to rapidly acquire advanced technologies, thus enhancing their innovation output \citep{ahujaTechnologicalAcquisitionsInnovation2001}. When firms face competition from foreign competitors, especially if these competitors have cost or technological advantages, local firms may seek to acquire advanced technologies through M\&As to enhance their competitiveness. However, technological M\&As are also accompanied by high costs and risks. The previous section explored the impact of tariff reduction on vertical integration, we further explore whether firms pursue vertical integration with the aim of acquiring technology under increased import competition.
	
	\subsubsection{Definition}\noindent
	Following \citet{ahujaTechnologicalAcquisitionsInnovation2001}, we define technological M\&As as: 1. If the description of an M\&A event explicitly mentions that the purpose is to obtain technology, we consider this M\&A event to be a technological M\&A. Specifically, the zephyr database has a ``deal comment'' for each M\&A event, which provides a comprehensive deal review for each transaction, and the product is described as follows:``Our researchers consolidated the information into a comprehensive deal review presented in chronological order. Additional information includes deal structure, financing and payment methods, and references to regulatory or shareholder issues. The commentary is updated as the deal progresses. The strategic rationale behind the transaction is highlighted through quotes when available". We determined whether the M\&A was a technological M\&A by manually identifying this content; 2. If the industry of the target company is a high-tech industry, we define this event as a technological M\&A. The definition of high-tech industry comes from National Bureau of Statistics (NBS), in the end, we identified a total of 7,867 technological M\&A events.
	
	\subsubsection{Result}\noindent
	We now empirically test whether import competition affects backward technological M\&As. Using the industries' upstreamness, we segregate backward technological M\&As from our overall sample, column (3) of \Cref{hightech} shows that 1\% decrease in downstream tariffs increases the probability of firms engaging in technological M\&As by 3.5\%, column (4) shows that tariff reduction doesn't have a significant effect on backward technological M\&As. Based on our previous analysis, high-tech firms may exhibit heterogeneity. We further examine whether technology is a potential mechanism through which import competition affects vertical integration in high-tech firms. As shown in columns (5)-(6) of \Cref{hightech}, tariff reductions promote backward technology-driven M\&As among high-tech firms, while the effect is insignificant for non-high-tech firms. Specifically, for every 1\% decrease in tariffs, the expected number of backward technology-driven acquisitions by high-tech firms increases by 13.6\%.\par
	Overall, the results in \Cref{hightech} indicates that high-tech firms are more likely to engage in vertical integration in response to tariff reduction, compared to non-high-tech firms. Furthermore, tariff reduction lead to more import competition, which bring more uncertainty and risks, although we find import competition promote firms to engage more in technology-driven M\&As, but as soon as we focus on vertical directions, this effect is only significant among high-tech firms.
	
	\section{Conclusion}\noindent
	Over the past two decades, trade liberalization has become the foreign trade policy of many developing countries, including China. Under this background, how will firms adjust their production organizational choices? We first build a Nash-bargaining model based on previous studies, the model predicts that a fall in downstream tariff induces more vertical integration; a fall in upstream tariff enhances this effect. since the effect is caused by reduce the investment of upstream, we predict that this effect is more pronounced in high asset specificity industries.\par
	We then select Chinese listed firms' data from 2000 to 2023 to examine our hypotheses. We use China's import tariff changes as an exogenous shock to represent import competition. Our empirical results show that tariff reduction has a significant positive effect on vertical integration and passes a series of robustness tests.\par
	Subsequently, we investigate the heterogeneous effects of tariffs on vertical integration and the underlying mechanisms. First, we categorize firms into state-owned enterprises (SOEs) and non-SOEs, we find that tariff reduction mainly promotes vertical integration in non-SOEs. Next, we group firms from low to high into four groups (Q1 to Q4) based on the industry's upstreamness. The impact of tariff reductions is most pronounced among firms in the Q1 and Q3. In our mechanism tests, we discover that controlling for upstream tariffs undermine both the magnitude and significance level of the downstream's coefficient, simultaneously, the coefficient for upstream tariff is also negative, indicating that upstream tariff enhances the effect of downstream tariff. Furthermore, we find that import competition has a significantly larger impact on differentiated industries compared to homogeneous ones. This suggests that relationship-specific investments serve as another crucial mechanism through which import competition affects vertical integration.\par
	We further examine whether import competition can encourage more technology-driven vertical integration. We first find tariff reduction has a greater effect on high-tech firms; second, we distinguish between technological M\&As and non-technological M\&As, find that tariff reduction will promote the number of technological M\&As in total, but has no significant effect on backward technological M\&As; lastly, we separately analyze high-technology and non-high-technology firms. We find that import competition can encourage more technology-driven vertical integration, but only among high-tech firms. Our study provides novel perspectives and empirical evidence on how firms optimize their production organization in response to globalization.
	\section*{Acknowledgment}
	Du thanks the financial support by the Graduate Innovation Fund of Wuhan Textile University [523058]. Shi acknowledge the financial support by the National Social Science Youth Fund of China [21CJY019].
	\section*{Funding}
	The work was supported by the National Social Science Youth Fund of China [21CJY019]; the Graduate Innovation Fund of Wuhan Textile University [523058].
	\bibliographystyle{abbrvnat}
	\bibliography{workingpaper-arxiv}

\begin{thebibliography}{28}
\providecommand{\natexlab}[1]{#1}
\providecommand{\url}[1]{\texttt{#1}}
\expandafter\ifx\csname urlstyle\endcsname\relax
  \providecommand{\doi}[1]{doi: #1}\else
  \providecommand{\doi}{doi: \begingroup \urlstyle{rm}\Url}\fi

\bibitem[Acemoglu et~al.(2009)Acemoglu, Johnson, and
  Mitton]{acemogluDeterminantsVerticalIntegration2009}
D.~Acemoglu, S.~Johnson, and T.~Mitton.
\newblock Determinants of {{Vertical Integration}}: {{Financial Development}}
  and {{Contracting Costs}}.
\newblock \emph{The Journal of Finance}, 64\penalty0 (3):\penalty0 1251--1290,
  2009.

\bibitem[Acemoglu et~al.(2010)Acemoglu, Aghion, Griffith, and
  Zilibotti]{acemogluVerticalIntegrationTechnology2010}
D.~Acemoglu, P.~Aghion, R.~Griffith, and F.~Zilibotti.
\newblock Vertical {{Integration}} and {{Technology}}: {{Theory}} and
  {{Evidence}}.
\newblock \emph{Journal of the European Economic Association}, 8\penalty0
  (5):\penalty0 989--1033, 2010.

\bibitem[Ahuja and Katila(2001)]{ahujaTechnologicalAcquisitionsInnovation2001}
G.~Ahuja and R.~Katila.
\newblock Technological acquisitions and the innovation performance of
  acquiring firms: A longitudinal study.
\newblock \emph{Strategic Management Journal}, 22\penalty0 (3):\penalty0
  197--220, 2001.

\bibitem[Alfaro et~al.(2016)Alfaro, Conconi, Fadinger, and
  Newman]{alfaroPricesDetermineVertical2016}
L.~Alfaro, P.~Conconi, H.~Fadinger, and A.~F. Newman.
\newblock Do {{Prices Determine Vertical Integration}}?
\newblock \emph{The Review of Economic Studies}, 83\penalty0 (3):\penalty0
  855--888, 2016.

\bibitem[Amiti and Khandelwal(2013)]{amitiImportCompetitionQuality2013}
M.~Amiti and A.~K. Khandelwal.
\newblock Import {{Competition}} and {{Quality Upgrading}}.
\newblock \emph{Review of Economics and Statistics}, 95\penalty0 (2):\penalty0
  476--490, 2013.

\bibitem[Bernard et~al.(2018)Bernard, Moxnes, and
  {Ulltveit-Moe}]{bernardTwoSidedHeterogeneityTrade2018}
A.~B. Bernard, A.~Moxnes, and K.~H. {Ulltveit-Moe}.
\newblock Two-{{Sided Heterogeneity}} and {{Trade}}.
\newblock \emph{The Review of Economics and Statistics}, 100\penalty0
  (3):\penalty0 424--439, 2018.

\bibitem[Brandt et~al.(2017)Brandt, Van~Biesebroeck, Wang, and
  Zhang]{brandtWTOAccessionPerformance2017}
L.~Brandt, J.~Van~Biesebroeck, L.~Wang, and Y.~Zhang.
\newblock {{WTO Accession}} and {{Performance}} of {{Chinese Manufacturing
  Firms}}.
\newblock \emph{American Economic Review}, 107\penalty0 (9):\penalty0
  2784--2820, 2017.

\bibitem[Cattaneo et~al.(2024)Cattaneo, Crump, Farrell, and
  Feng]{cattaneoBinscatter2024}
M.~D. Cattaneo, R.~K. Crump, M.~H. Farrell, and Y.~Feng.
\newblock On {{Binscatter}}.
\newblock \emph{American Economic Review}, 114\penalty0 (5):\penalty0
  1488--1514, 2024.

\bibitem[Chor et~al.(2021)Chor, Manova, and Yu]{chorGrowingChinaFirm2021}
D.~Chor, K.~Manova, and Z.~Yu.
\newblock Growing like {{China}}: {{Firm}} performance and global production
  line position.
\newblock \emph{Journal of International Economics}, 130:\penalty0 103445,
  2021.

\bibitem[Chun et~al.(2020)Chun, Hur, Kim, and
  Son]{chunCrossBorderVerticalIntegration2020}
H.~Chun, J.~Hur, D.~Kim, and N.~S. Son.
\newblock Cross-{{Border Vertical Integration}} and {{Technology}} in {{Factory
  Asia}}: {{Evidence}} from {{Korea}}.
\newblock \emph{The Developing Economies}, 58\penalty0 (2):\penalty0 99--133,
  2020.

\bibitem[Costinot et~al.(2013)Costinot, Vogel, and
  Wang]{costinotElementaryTheoryGlobal2013}
A.~Costinot, J.~Vogel, and S.~Wang.
\newblock An {{Elementary Theory}} of {{Global Supply Chains}}.
\newblock \emph{The Review of Economic Studies}, 80\penalty0 (1):\penalty0
  109--144, 2013.

\bibitem[Fan et~al.(2017)Fan, Huang, Morck, and
  Yeung]{fanInstitutionalDeterminantsVertical2017}
J.~P. Fan, J.~Huang, R.~Morck, and B.~Yeung.
\newblock Institutional determinants of vertical integration in {{China}}.
\newblock \emph{Journal of Corporate Finance}, 44:\penalty0 524--539, 2017.

\bibitem[Fernandes et~al.(2022)Fernandes, Kee, and
  Winkler]{fernandesDeterminantsGlobalValue2022}
A.~M. Fernandes, H.~L. Kee, and D.~Winkler.
\newblock Determinants of {{Global Value Chain Participation}}: {{Cross-Country
  Evidence}}.
\newblock \emph{The World Bank Economic Review}, 36\penalty0 (2):\penalty0
  329--360, 2022.

\bibitem[Gereffi et~al.(2005)Gereffi, Humphrey, and
  Sturgeon]{gereffiGovernanceGlobalValue2005}
G.~Gereffi, J.~Humphrey, and T.~Sturgeon.
\newblock The governance of global value chains.
\newblock \emph{Review of International Political Economy}, 12\penalty0
  (1):\penalty0 78--104, 2005.

\bibitem[Grossman and Hart(1986)]{grossmanCostsBenefitsOwnership1986}
S.~J. Grossman and O.~D. Hart.
\newblock The {{Costs}} and {{Benefits}} of {{Ownership}}: {{A Theory}} of
  {{Vertical}} and {{Lateral Integration}}.
\newblock \emph{Journal of Political Economy}, 94\penalty0 (4):\penalty0
  691--719, 1986.

\bibitem[Hart and Moore(1990)]{hartPropertyRightsNature1990}
O.~Hart and J.~Moore.
\newblock Property {{Rights}} and the {{Nature}} of the {{Firm}}.
\newblock \emph{Journal of Political Economy}, 98\penalty0 (6):\penalty0
  1119--1158, 1990.

\bibitem[Liu and Qiu(2016)]{liuIntermediateInputImports2016}
Q.~Liu and L.~D. Qiu.
\newblock Intermediate input imports and innovations: {{Evidence}} from
  {{Chinese}} firms' patent filings.
\newblock \emph{Journal of International Economics}, 103:\penalty0 166--183,
  2016.

\bibitem[Liu et~al.(2019)Liu, Qiu, and
  Zhan]{liuTradeLiberalizationDomestic2019}
Q.~Liu, L.~D. Qiu, and C.~Zhan.
\newblock Trade liberalization and domestic vertical integration: {{Evidence}}
  from {{China}}.
\newblock \emph{Journal of International Economics}, 121:\penalty0 103250,
  2019.

\bibitem[Liu et~al.(2021)Liu, Lu, Lu, and Luong]{liuImportCompetitionFirm2021}
Q.~Liu, R.~Lu, Y.~Lu, and T.~A. Luong.
\newblock Import competition and firm innovation: {{Evidence}} from {{China}}.
\newblock \emph{Journal of Development Economics}, 151:\penalty0 102650, 2021.

\bibitem[Lu and Yu(2015)]{luTradeLiberalizationMarkup2015}
Y.~Lu and L.~Yu.
\newblock Trade {{Liberalization}} and {{Markup Dispersion}}: {{Evidence}} from
  {{China}}'s {{WTO Accession}}.
\newblock \emph{American Economic Journal: Applied Economics}, 7\penalty0
  (4):\penalty0 221--253, 2015.

\bibitem[McLaren(2000)]{mclarenGlobalizationVerticalStructure2000}
J.~McLaren.
\newblock "{{Globalization}}" and {{Vertical Structure}}.
\newblock \emph{American Economic Review}, 90\penalty0 (5):\penalty0
  1239--1254, 2000.

\bibitem[Ornelas and Turner(2008)]{ornelasTradeLiberalizationOutsourcing2008a}
E.~Ornelas and J.~L. Turner.
\newblock Trade liberalization, outsourcing, and the hold-up problem.
\newblock \emph{Journal of International Economics}, 74\penalty0 (1):\penalty0
  225--241, 2008.

\bibitem[Ornelas and Turner(2012)]{ornelasProtectionInternationalSourcing2012}
E.~Ornelas and J.~L. Turner.
\newblock Protection and {{International Sourcing}}.
\newblock \emph{The Economic Journal}, 122\penalty0 (559):\penalty0 26--63,
  2012.

\bibitem[Qiu and Zhou(2013)]{qiuMultiproductFirmsScope2013}
L.~D. Qiu and W.~Zhou.
\newblock Multiproduct firms and scope adjustment in globalization.
\newblock \emph{Journal of International Economics}, 91\penalty0 (1):\penalty0
  142--153, 2013.

\bibitem[Rauch(1999)]{rauchNetworksMarketsInternational1999}
J.~E. Rauch.
\newblock Networks versus markets in international trade.
\newblock \emph{Journal of International Economics}, 48\penalty0 (1):\penalty0
  7--35, 1999.

\bibitem[Stiebale and Vencappa(2022)]{stiebaleImportCompetitionVertical2022}
J.~Stiebale and D.~Vencappa.
\newblock Import competition and vertical integration: {{Evidence}} from
  {{India}}.
\newblock \emph{Journal of Development Economics}, 155:\penalty0 102790, 2022.

\bibitem[Topalova and Khandelwal(2011)]{topalovaTradeLiberalizationFirm2011}
P.~Topalova and A.~Khandelwal.
\newblock Trade {{Liberalization}} and {{Firm Productivity}}: {{The Case}} of
  {{India}}.
\newblock \emph{Review of Economics and Statistics}, 93\penalty0 (3):\penalty0
  995--1009, 2011.

\bibitem[Williamson(1979)]{williamsonTransactionCostEconomicsGovernance1979}
O.~E. Williamson.
\newblock Transaction-{{Cost Economics}}: {{The Governance}} of {{Contractual
  Relations}}.
\newblock \emph{The Journal of Law and Economics}, 22\penalty0 (2):\penalty0
  233--261, 1979.

\end{thebibliography}

\begin{appendices}
	\clearpage
\section{Tables}
\captionsetup{justification=raggedright, singlelinecheck=false,skip=0pt,labelsep=space}
\begin{table}[htbp]
	\centering
	\caption{\\Control variables}
	\label{control}
	\resizebox{\textwidth}{!}{%
		\begin{threeparttable}
			\begin{tabularx}{\textwidth}{l*{1}{X}}
				\toprule
				{\textbf{Variables}} & {\textbf{Calculation method}} \\ \midrule
				firm age & ln(firm age + 1), where firm age is calculated as the difference between the current year and the firm's year of establishment \\
				firm size & ln(firm revenue + 1) \\
				skill density & ln(average wage per employee) \\
				liquidity & ln(current ratio), where CurrentRatio is defined as current assets divided by current liabilities \\
				leverage ratio & ln(asset-to-liability ratio), where AssetLiabilityRatio is defined as total assets divided by total liabilities \\
				HHI index & $\ln({\sum_{i}\frac{turnover_i}{turnover_{total}}})$ \\ \bottomrule
				\multicolumn{2}{l}{Note: Data calculated and compiled by the authors}
			\end{tabularx}
		\end{threeparttable}
	}
\end{table}
\begin{table}[htbp]
	\centering
	\captionsetup{justification=raggedright, singlelinecheck=false,skip=0pt,labelsep=space}
	\caption{\\Descriptive statistics}
	\label{describe}
	\begin{threeparttable}
		\begin{tabularx}{\textwidth}{l*{5}{X}}
			\toprule
			& (1) & (2) & (3) & (4) & (5) \\ 
			VARIABLES & N & mean & sd & min & max \\ \midrule
			\multicolumn{6}{l}{\textit{panel A: key variables}} \\ \cmidrule(lr){1-2}
			downstream\_tariff & 52,079 & 9.502 & 5.500 & 0 & 30 \\
			upstream\_tariff & 52,446 & 5.061 & 1.912 & 0.929 & 19.41 \\
			ahs\_tariff & 51,819 & 8.949 & 5.507 & 0 & 30 \\
			ind3\_tariff & 51,937 & 9.456 & 4.478 & 0 & 25 \\
			weight\_tariff & 48,637 & 8.237 & 6.388 & 0 & 31.82 \\
			backward & 52,474 & 0.0583 & 0.234 & 0 & 1 \\
			backward\_num & 52,474 & 0.0789 & 0.402 & 0 & 18 \\
			\multicolumn{6}{l}{\textit{panel B: binary variables}} \\ \cmidrule(lr){1-2}
			high\_tech & 52,474 & 0.188 & 0.391 & 0 & 1 \\
			soe & 52,474 & 0.287 & 0.452 & 0 & 1 \\
			isic\_lib & 52,474 & 0.665 & 0.472 & 0 & 1 \\
			\multicolumn{6}{l}{\textit{panel C: control variables}} \\ \cmidrule(lr){1-2}
			age & 52,474 & 2.445 & 0.756 & 0 & 4.190 \\
			size & 36,772 & 14.04 & 1.367 & 10.85 & 18.04 \\
			leverage & 36,651 & 0.350 & 0.131 & 0.0581 & 0.663 \\
			liquidity & 36,590 & 1.041 & 0.430 & 0.282 & 2.726 \\
			skill & 35,966 & 11.74 & 1.327 & 8.521 & 15.42 \\
			hhi & 52,373 & 0.196 & 0.193 & 0 & 1 \\ \bottomrule
			\multicolumn{6}{l}{Note: Data calculated and compiled by the authors}
		\end{tabularx}
	\end{threeparttable}
\end{table}
\begin{table}[htbp]
	\centering
	\caption{\\Baseline Result}
	\label{baseline}
	\begin{threeparttable}
		\begin{tabularx}{\textwidth}{l*{4}{X}}
			\toprule
			& (1) & (2) & (3) & (4) \\ \midrule
			downstream\_tariff & -0.024*** & -0.050*** & -0.014*** & -0.044*** \\
			& (-4.96) & (-3.88) & (-3.09) & (-3.26) \\
			age &  &  & 0.593*** & 2.267*** \\
			&  &  & (9.70) & (6.54) \\
			size &  &  & 0.088** & 0.253*** \\
			&  &  & (2.23) & (3.73) \\
			skill &  &  & 0.331*** & 0.404*** \\
			&  &  & (7.65) & (4.97) \\
			leverage &  &  & -0.168 & -1.041** \\
			&  &  & (-0.46) & (-2.23) \\
			liquidity &  &  & 0.150 & -0.005 \\
			&  &  & (1.54) & (-0.04) \\
			hhi &  &  & -0.018 & -1.016** \\
			&  &  & (-0.12) & (-1.99) \\
			Constant & -2.058*** & -0.696*** & -9.196*** & -15.569*** \\
			& (-41.31) & (-6.20) & (-27.00) & (-12.40) \\
			Observations & 34,409 & 18,039 & 34,409 & 18,039 \\
			company   FE & NO & YES & NO & YES \\
			industry   FE & NO & YES & NO & YES \\
			year   FE & NO & YES & NO & YES \\
			Pseudo   R-squared & 0.00219 & 0.174 & 0.0640 & 0.191 \\ \bottomrule
			\multicolumn{5}{l}{Note: Robust standard error clustered at firm level. Robust z-statistics in parentheses.} \\
			\multicolumn{5}{l}{*** p$<$.01, ** p$<$.05, * p$<$.1} \\
		\end{tabularx}
	\end{threeparttable}
\end{table}
\begin{table}[htbp]
	\centering
	\caption{\\Robustness checks 1}
	\label{robust1}
	\begin{threeparttable}
		\resizebox{\textwidth}{!}{%
			\begin{tabular}{lllllll}
				\toprule
				& (1) & (2) & (3) & (4) & (5) & (6) \\ \cmidrule(lr){2-2} \cmidrule(lr){3-3} \cmidrule(lr){4-4} \cmidrule(lr){5-5} \cmidrule(lr){6-6} \cmidrule(lr){7-7} 
				& OLS & Logit & leverage$<$1 & manufacturer & financial crisis & All three \\ \midrule
				downstream\_tariff & -0.004*** & -0.041*** & -0.044*** & -0.042*** & -0.044*** & -0.042*** \\
				& (-2.98) & (-2.87) & (-3.26) & (-3.09) & (-3.17) & (-3.01) \\
				Constant & -0.872*** &  & -15.569*** & -15.309*** & -16.243*** & -16.004*** \\
				& (-11.19) &  & (-12.40) & (-12.04) & (-11.72) & (-11.44) \\
				Observations & 34,407 & 18,868 & 18,039 & 17,442 & 16,257 & 15,709 \\
				controls & YES & YES & YES & YES & YES & YES \\
				company   FE & YES & YES & YES & YES & YES & YES \\
				industry   FE & YES & YES & YES & YES & YES & YES \\
				year   FE & YES & YES & YES & YES & YES & YES \\
				Pseudo   R-squared &  &  & 0.191 & 0.191 & 0.187 & 0.187\\ \bottomrule
				\multicolumn{7}{l}{Note: Robust standard error clustered at firm level. Robust z-statistics in parentheses.} \\
				\multicolumn{7}{l}{*** p$<$.01, ** p$<$.05, * p$<$.1}
			\end{tabular}
		}
	\end{threeparttable}
\end{table}
\begin{table}[htbp]
	\centering
	\caption{\\Robustness checks 2}
	\label{robust2}
	\resizebox{\textwidth}{!}{%
		\begin{threeparttable}
			\begin{tabular}{llllll}
				\toprule
				& (1) & (2) & (3) & (4) & (5) \\ \cmidrule(lr){2-2} \cmidrule(lr){3-3} \cmidrule(lr){4-4} \cmidrule(lr){5-5} \cmidrule(lr){6-6}
				& ahs\_tariff & 3digit-industry & weight\_tariff & lag-one & extra-control \\ \midrule
				downstream\_tariff & -0.026** & -0.041*** & -0.018* & -0.041*** & -0.035** \\
				& (-2.04) & (-2.66) & (-1.86) & (-3.27) & (-2.22) \\
				growth &  &  &  &  & -0.036 \\
				&  &  &  &  & (-0.88) \\
				tangibility &  &  &  &  & -1.490*** \\
				&  &  &  &  & (-3.38) \\
				k\_density &  &  &  &  & 0.274*** \\
				&  &  &  &  & (3.44) \\
				Constant & -15.848*** & -15.682*** & -16.062*** & -9.312*** & -13.379*** \\
				& (-12.48) & (-12.43) & (-12.31) & (-10.94) & (-7.77) \\
				Observations & 17,891 & 18,005 & 17,264 & 19,194 & 10,667 \\
				controls & YES & YES & YES & YES & YES \\
				company FE & YES & YES & YES & YES & YES \\
				industry FE & YES & YES & YES & YES & YES \\
				year FE & YES & YES & YES & YES & YES \\
				Pseudo R-squared & 0.191 & 0.191 & 0.189 & 0.170 & 0.187 \\ \bottomrule
				\multicolumn{6}{l}{Note: Robust standard error clustered at firm level. Robust z-statistics in parentheses.} \\
				\multicolumn{6}{l}{*** p$<$.01, ** p$<$.05, * p$<$.1}
			\end{tabular}
		\end{threeparttable}
	}
\end{table}
\begin{table}[htbp]
	\centering
	\caption{\\Heterogeneous effects}
	\label{hetero1}
	\resizebox{\textwidth}{!}{
		\begin{threeparttable}
			\begin{tabular}{lllllll}
				\toprule
				& (1) & (2) & (3) & (4) & (5) & (6) \\
				& \multicolumn{2}{l}{backward} & \multicolumn{4}{l}{upstreamness, from low to high} \\ \cmidrule(lr){2-3} \cmidrule(lr){4-7}
				& Non-soe & soe & Q1 & Q2 & Q3 & Q4 \\ \midrule
				downstream\_tariff & -0.046*** & -0.037 & -0.103*** & -0.020 & -0.084** & -0.050 \\
				& (-3.02) & (-1.28) & (-2.62) & (-1.11) & (-2.54) & (-1.05) \\
				Constant & -15.097*** & -15.827*** & -13.440*** & -16.125*** & -15.476*** & -16.311*** \\
				& (-9.20) & (-7.44) & (-5.43) & (-7.71) & (-6.44) & (-5.45) \\
				Observations & 11,510 & 6,293 & 5,400 & 5,129 & 3,542 & 3,516 \\
				controls & YES & YES & YES & YES & YES & YES \\
				company   FE & YES & YES & YES & YES & YES & YES \\
				industry   FE & YES & YES & YES & YES & YES & YES \\
				year   FE & YES & YES & YES & YES & YES & YES \\
				Pseudo   R-squared & 0.179 & 0.210 & 0.194 & 0.178 & 0.206 & 0.183 \\ \bottomrule
				\multicolumn{7}{l}{Note: Robust standard error clustered at firm level. Robust z-statistics in parentheses.} \\
				\multicolumn{7}{l}{*** p$<$.01, ** p$<$.05, * p$<$.1}
			\end{tabular}
		\end{threeparttable}
	}
\end{table}
\begin{table}[htbp]
	\centering
	\caption{\\Mechanism tests}
	\label{mechanism}
	\resizebox{\textwidth}{!}{%
		\begin{threeparttable}
			\begin{tabularx}{\textwidth}{l*{4}{X}}
				\toprule
				& (1) & (2) & (3) & (4) \\ \cmidrule(lr){2-3} \cmidrule(lr){4-4} \cmidrule(lr){5-5}
				& \multicolumn{2}{l}{Upstream tariff} & Homo & Different \\ \midrule
				downstream\_tariff & -0.045*** & -0.034** & -0.040 & -0.041*** \\
				& (-3.29) & (-2.21) & (-1.25) & (-2.72) \\
				upstream\_tariff &  & -0.188* &  &  \\
				&  & (-1.67) &  &  \\
				Constant & -15.571*** & -14.772*** & -13.835*** & -15.943*** \\
				& (-12.41) & (-11.17) & (-6.59) & (-10.02) \\
				Observations & 18,025 & 18,025 & 5,843 & 12,076 \\
				controls & YES & YES & YES & YES \\
				company FE & YES & YES & YES & YES \\
				industry FE & YES & YES & YES & YES \\
				year FE & YES & YES & YES & YES \\
				Pseudo R-squared & 0.191 & 0.192 & 0.184 & 0.199 \\ \bottomrule
				\multicolumn{5}{l}{Note: Robust standard error clustered at firm level. Robust z-statistics in parentheses.} \\
				\multicolumn{5}{l}{*** p$<$.01, ** p$<$.05, * p$<$.1}
			\end{tabularx}
		\end{threeparttable}
	}
\end{table}
\begin{table}[htbp]
	\caption{\\Further discussion}
	\label{hightech}
	\resizebox{\textwidth}{!}{%
		\begin{threeparttable}
			\begin{tabular}{lllllll}
				\toprule
				& (1) & (2) & (3) & (4) & (5) & (6) \\
				& \multicolumn{2}{l}{backward} & \multicolumn{2}{l}{technical} & \multicolumn{2}{l}{tech\_backward} \\ \cmidrule(lr){2-3} \cmidrule(lr){4-5} \cmidrule(lr){6-7}
				& Non & High-tech & ma & backward & Non & High \\ \midrule
				downstream\_tariff & -0.035** & -0.092** & -0.035*** & -0.026 & 0.001 & -0.136* \\
				& (-2.49) & (-2.37) & (-2.92) & (-1.31) & (0.05) & (-1.81) \\
				Constant & -16.004*** & -13.736*** & -0.951*** & -1.381*** & -13.647*** & -18.832*** \\
				& (-12.80) & (-3.78) & (-9.78) & (-8.48) & (-5.63) & (-3.18) \\
				Observations & 14,882 & 2,808 & 27,424 & 8,573 & 5,032 & 1,054 \\
				controls & YES & YES & YES & YES & YES & YES \\
				company   FE & YES & YES & YES & YES & YES & YES \\
				industry   FE & YES & YES & YES & YES & YES & YES \\
				year FE & YES & YES & YES & YES & YES & YES \\
				Pseudo   R-squared & 0.189 & 0.191 & 0.216 & 0.166 & 0.162 & 0.164 \\ \bottomrule
				\multicolumn{7}{l}{Note: Robust standard error clustered at firm level. Robust z-statistics in parentheses.} \\
				\multicolumn{7}{l}{*** p$<$.01, ** p$<$.05, * p$<$.1}
			\end{tabular}
		\end{threeparttable}
	}
\end{table}
\clearpage

\section{Figures}
\begin{figure}
	\centering
	\captionsetup{justification=centering,skip=0pt}
	\includegraphics[width=0.8\textwidth]{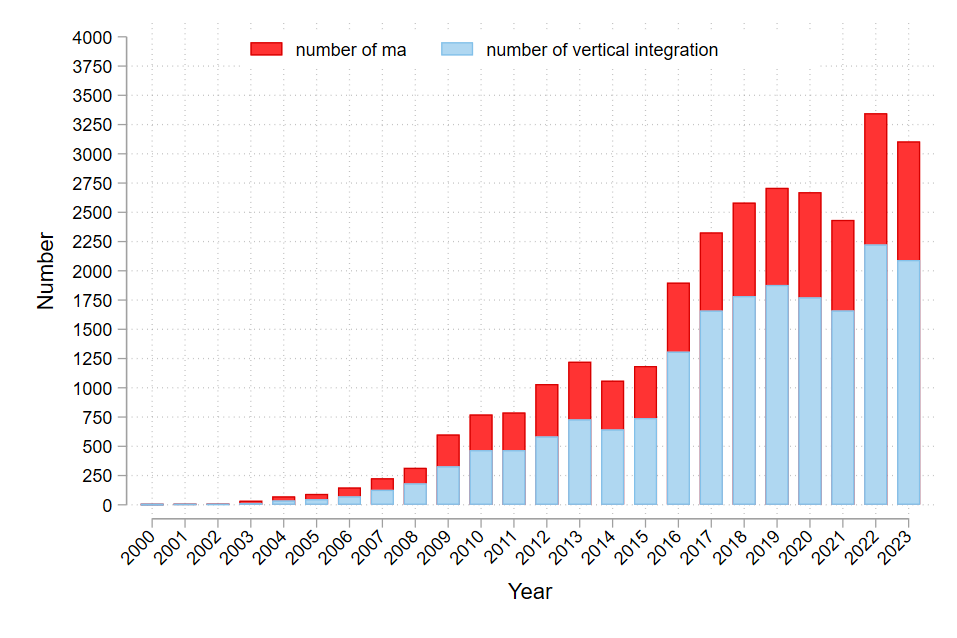}
	\caption{Domestic M\&As in China between 2000-2023}
	\label{fig_ma}
\end{figure}
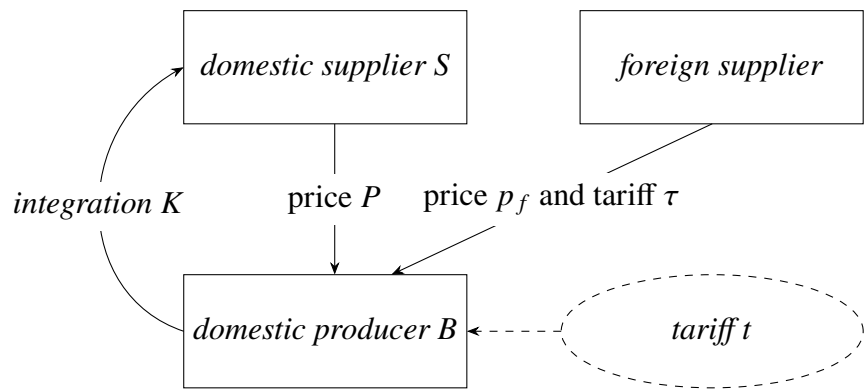
\begin{figure}[!ht]
	\centering
	\captionsetup{justification=centering,skip=20pt}
	\resizebox{0.70\textwidth}{!}{%
		\begin{circuitikz}
			\tikzstyle{every node}=[font=\normalsize]

			\draw  (2,15.75) rectangle  node {\normalsize \textit{domestic supplier S}} (5.75,14.25);
			\draw  (7.25,15.75) rectangle  node {\normalsize \textit{foreign supplier}} (11,14.25);
			\draw  (2,12.25) rectangle  node {\normalsize \textit{domestic producer B}} (5.75,10.75);
			\draw [->, >=Stealth] (4,14.25) -- (4,12.25)node[pos=0.5, fill=white]{price $P$};
			\draw [->, >=Stealth] (9,14.25) -- (4.75,12.25)node[pos=0.5, fill=white]{price $p_f$ and tariff $\tau$};
			\draw [ dashed] (9,11.5) ellipse (2cm and 0.75cm) node {\normalsize \textit{tariff $t$}} ;
			\draw [->, >=Stealth, dashed] (7,11.5) -- (5.75,11.5);
			
			\draw [->, >=Stealth] (2,11.5) .. controls (0.5,12) and (0.5,14.25) .. (2,15) node[pos=0.5, fill=white]{\textit{integration $K$}};
		\end{circuitikz}
	}%
	\caption{The connection between $S$ and $B$}
	\label{relation}
\end{figure}
\end{appendices}
\end{document}